\documentstyle[pra,aps,eqsecnum]{revtex}
\begin{document}
\def\be{\begin{eqnarray}}
\def\ee{\end{eqnarray}}
\def\l{\langle}
\def\r{\rangle}
\title
{
Reconstruction of quantum states of spin systems via the
Jaynes principle of maximum entropy}

\author{V. Bu\v{z}ek$^{1,2}$, G. Drobn\'{y}$^{2,3}$,  G. Adam$^{4}$,
R. Derka$^{5}$,  and P.L. Knight$^{1}$
}
\address{
$^{1}$ Optics Section, The Blackett Laboratory,
Imperial College, London SW7 2BZ,  England\\
$^{2}$ Institute of Physics, Slovak Academy of Sciences,
D\'ubravsk\'a cesta 9, 842 28 Bratislava, Slovakia\\
$^{3}$ Arbeitsgruppe ``Nichtklassische Strahlung'', MPG,
 Rudower Chaussee 5, 12484 Berlin, Germany\\
$^{4}$ Institut f\"ur Theoretische Physik, Technische Universit\"at Wien,
    Wiedner Hauptstrasse 8-10, A-1040 Vienna, Austria\\
$^{5}$ Department of Physics, Oxford University, Parks Road,
OX1 3PU Oxford, England
}

\date{November 12, 1996}
\maketitle

\begin{abstract}
We apply the Jaynes principle of  {\em maximum entropy}
for the partial reconstruction of correlated spin states.
We determine
the minimum set of observables which are necessary
for the complete reconstruction of
the most correlated states of systems composed of
spins--1/2  (e.g., the Bell and the Greenberger--Horne--Zeilinger
states).
We investigate to what extent an incomplete measurement can
reveal nonclassical features of correlated spin states.

\end{abstract}
\pacs{03.65.Bz}


\section{INTRODUCTION}

The seminal paper by Vogel and Risken \cite{Vogel89} on the tomographic
reconstruction of Wigner functions of light fields has greatly enhanced
interest into the old problem of ``measurement'' of states of
quantum-mechanical systems. Within the last few years tomographic
reconstruction has been experimentally realized for example
by Raymer and coworkers
\cite{Smithey93}, and Mlynek and coworkers \cite{Mlynek}.
Tomographic reconstruction schemes of states
of other bosonic systems such as vibrational modes of trapped
atoms \cite{Vogel} and atomic waves \cite{Wilkens} have
been proposed. Recently Wigner functions of  vibration states
of a trapped atom have been experimentally determined by Wineland
and coworkers \cite{Leibfried}, while Kurtsiefer and coworkers
\cite{Kurtsiefer}
have measured Wigner functions of atomic wave packets.
 Leonhardt \cite{Leonhardt} has extended the ideas
of Vogel and Risken to the case of Wigner functions in discrete
phase spaces associated with physical systems with finite-dimensional
Hilbert spaces, such as spin systems.

The problem of
reconstruction of states of finite-dimensional systems is closely
related to various aspects of quantum information processing,
such as
 reading of  registers of quantum computers
\cite{Barenco}. This problem also emerges when states
of atoms are reconstructed. In particular, Walser, Cirac,
and Zoller \cite{Walser} have shown that under certain conditions
quantum states of a single quantized cavity mode can be
{\em completely} transferred on to the internal Zeeman submanifold of an atom.
Consequently, the reconstruction of the states of a cavity mode is reduced
to the problem of reconstruction of angular momentum states in a finite
dimensional Hilbert space.

From the  postulates of quantum mechanics it follows that the
{\em complete} reconstruction of a state of a quantum-mechanical system
can be performed providing a complete set of system observables
(i.e., the {\em quorum} \cite{Band})
is measured
on the ensemble of identically prepared systems. This goal
on one hand may be technically difficult to realize and on the other hand
it may not be necessary. In many situations even partial knowledge
(i.e., incomplete reconstruction) of the state is sufficient
for particular purposes.

The complete reconstruction of spin states have been addressed
in the literature \cite{Leonhardt,Newton,Band,Wootters}.
In the present paper we will analyze  the problem of the partial
reconstruction
of these states. We will show how the spin state can be reconstructed
when just a restricted set of mean values of the system observables
is known from the measurement. Utilizing the Jaynes principle of
{\em maximum entropy} we will partially  reconstruct  the density
operator from the available (i.e., measured) mean values of system
observables.

We will also address the question as to which is the minimum observation level
(i.e., a specific
subset of system observables) on which the complete reconstruction
can be	performed. In particular, we will analyze the reconstruction of the
most correlated states of the system composed of two and three spins-1/2
(i.e., the Bell and the Greenberger--Horne--Zeilinger  states).

The present paper is organized as follows. In Section II we briefly review
the principle of maximum entropy and the formalism of
state reconstruction associated with
particular observation levels. In Section III we present a simple
illustration of a reconstruction of a state of a single spin-1/2.
Section IV will be devoted to the detailed analysis of the state reconstruction
for a system of two correlated spins-1/2.
In Section V we will address the problem
of the (partial) reconstruction of the Greenberger--Horne--Zeilinger states.
In Appendix A we present detailed reconstruction of density operators
on two nontrivial observation levels.

\section{RECONSTRUCTION OF DENSITY OPERATORS OF QUANTUM STATES}

When it is {\em a priori} known that
experimental data contain the complete information about the
state of the system, then it is just a question of technical
convenience how to perform a transformation of this data
into a more familiar object such a density operator. A particular
example of this procedure is  quantum homodyne tomography
\cite{Vogel89} when from the measured probability distributions
of rotated quadratures one can reconstruct\footnote{Strictly
speaking, Wigner functions or wave vectors cannot be measured,
they only can be reconstructed from experimental data with a help
of some inversion procedures.}
(with the help of the
inverse Radon transformation) the  Wigner function of the
state.

Now we can ask the question: ``What is the density operator of the
quantum mechanical system when an {\em incomplete} measurement	over
this system is performed?\,''. In this case the experimental data does not
provide us with sufficient information to specify the density
operator of the system uniquely, i.e. there can be many density operators
which fulfill the constraints imposed by incomplete experimental data.
In this situation one can only {\em estimate} what is the most probable
density operator which describes the system.

In principle we can distinguish two different forms of incompleteness:
firstly,  when a {\em precise} knowledge of a {\em subset} of
the quorum \cite{Band}
of system observables is known; secondly, when system observables are
not measured precisely,
i.e., instead of probability distributions only frequencies of appearances
of eigenvalues of these observables are available.

In the present paper we will focus our attention on the reconstruction
of density operators of spin states when mean values of a subset of system
observables are measured precisely. In this case the estimation of
density operators can be performed with the help of Jaynes principle
of maximum entropy.

\subsection{ Principle of maximum entropy and observation levels}
Let us assume that the state of a spin system,
described by the density operator
$\hat\rho_0$,
is unknown and only expectation values $G_\nu$ of observables
$\hat G_\nu$ ($\nu = 1,\ldots,n$) are available from a measurement.
The set of observables is referred to as the {\em observation level}
${\cal O}$ \cite{Fick}.
There can be a large number of density operators
$\hat\rho$ which are in agreement with the experimental results, i.e.,
\be
\mbox{Tr} (\hat\rho\hat{G}_\nu )=G_\nu \qquad (\nu = 1,\ldots,n)
\label{1.1}
.\ee
If we wish to use only the expectation values $G_\nu$ of the chosen
observation level for an estimation (reconstruction)
  of the density operator, then we face the problem of selecting
one particular density operator$\hat\rho _{\cal O}$   out of many
$\hat\rho$ which fulfill condition  (\ref{1.1}).
To perform this ``selection'' (i.e., estimation) we note that
the density operators under consideration
do differ by their degree of deviation from pure states.
To quantify this deviation an uncertainty measure has to be introduced.
Following Jaynes \cite{Jaynes} one can utilize
the von Neumann entropy \cite{vonNeumann}
\be
S[\hat\rho]=-\mbox{Tr}(\hat\rho \ln \hat\rho )
.\label{1.3} \ee
For pure states $S=0$ while for statistical mixtures of pure states
$S>0$.

According to the Jaynes principle of  maximum entropy,
we have to choose from a set of density operators $\hat\rho$
which fulfill the constraints of Eq.(\ref{1.1})
the generalized canonical density
operator $\hat\rho_{\cal O}$ which maximizes the value
of the von Neumann entropy\footnote{
We note that, in principle, instead of the von Neumann entropy one can
utilize another uncertainty measure to distinguish between the density
operators which fulfill constraints (\ref{1.1}).
For example, by maximizing the linearized entropy \cite{Wehrl}
\be
\nonumber
\eta[\hat\rho]=1-\mbox{Tr} (\hat\rho^2)
\ee
one can obtain a {\it partially} reconstructed density operator
$\hat\rho_{\cal O}^{(m)}$ which fulfills conditions (\ref{1.1}) and
simultaneously	leads to the ``maximum mixture'',
i.e., $\eta[\rho_{\cal O}^{(m)}]=\max$.
Note that for pure states $\eta=0$ while for mixtures $\eta>0$.
On the complete observation level, this ``maximum-mixture'' principle is
equivalent to the maximum-entropy principle but, in general, on the reduced
(incomplete) observation levels,
$\hat\rho_{\cal O}\neq\hat\rho_{\cal O}^{(m)}$.}.
In other words, the maximum-entropy principle is {\em the most conservative}
assignment, in the sense that it does not permit one to draw any
conclusions unwarranted by  the measured data.
The generalized canonical density operator $\hat\rho_{\cal O}$
represents a {\em partially} reconstructed (estimated)
density operator
on the given observation level ${\cal O}$.
The corresponding entropy $S_{\cal O}=S[\hat\rho_{\cal O}]$
represents the measure of deviation of the reconstructed state
from an original pure state.
The generalized canonical density operator $\hat\rho_{\cal O}$
takes the form \cite{Fick}
\be
\hat\rho_{\cal O}&=&{1 \over Z_{\cal O}}
\exp \left(-\sum\nolimits_\nu \lambda_\nu \hat{G}_\nu \right),
\label{1.4} \\
Z_{\cal O}&=& \mbox{Tr}
\left[\exp \left(-\sum\nolimits_\nu \lambda_\nu \hat{G}_\nu \right)
\right]
\nonumber
\ee
where $Z_{\cal O}$ is the generalized partition function;
$\lambda_\nu$ are the Lagrange multipliers which have to be found
from the set of equations  (\ref{1.1}).

Any incomplete observation level ${\cal O}_A$ can be extended
to a more complete observation level ${\cal O}_B$
which includes additional observables, i.e., ${\cal O}_A\subset{\cal O}_B$.
Additional expectation values
can only increase the amount of available information about the state of the
system. This procedure is called the {\em extension} of the observation
level (from ${\cal O}_A$  to ${\cal O}_B$) and is usually associated with
a decrease of the entropy, as $S_B \le S_A$.
We can also consider a {\em reduction} of the observation level if we
decrease number of independent observables which are measured.
This reduction is accompanied with an increase of the entropy due
to the decrease of information available about the state of the
system. Each incomplete observation level can be considered as a
reduction of the complete observation level.
In what follows we will study a sequence of observation levels in the form
\be
\begin{array}{ccccccccc}
  {\cal O}_A & \subset &{\cal O}_B & \subset & {\cal O}_C &
   \subset   & \ldots  & \subset   & {\cal O}_{\rm comp} \\
   S_A	     & \ge     & S_B	   &  \ge    &	S_C	   &
   \ge	     & \ldots  & \ge	   &  0~~~~
\end{array}
\label{1.5}
\ee
which represent successive extensions of an observation level
${\cal O}_A$ towards the complete observation level ${\cal O}_{\rm comp}$.

Concluding this section we make two remarks. \newline
{\bf (1)} Firstly
we stress that the reconstruction scheme
based on the Jaynes principle of maximum entropy does not
require any a priori assumption about the purity of  reconstructed
states, i.e., it can be applied for reconstruction of pure states
 as well as  for statistical  mixtures. This reconstruction
scheme is equivalent to an averaging over the generalized grand
canonical ensemble of {\em all} states of the system,
under the conditions imposed by the constraints given
by Eq.(\ref{1.1}). Within the framework  of a geometrical formalism,
each state of the quantum system is represented by a point
in the parametric state space\footnote{Pure states, which are elements
of the generalized microcanonical ensemble, are represented by points
on a manifold in this state space (such as the Poincare sphere
in the case of the spin-1/2).}. Those states which fulfils the
constraints (\ref{1.1}) are represented
by a specific manifold in the parametric space.
From the MaxEnt principle it then follows
that the generalized canonical density operator is equal to the
equally weighted average over all states on this specific manifold.
Obviously this average is represented by one special point which
is associated with the generalized canonical density operator.
\newline
{\bf (2)}
In the case when there is no information available about the
preparation of the system, then there is no intrinsic
way  to specify the ``minimal'' complete observation level.
Here by minimal we mean the complete observation level composed of the
smallest number of observables.  What one can do
is to extend systematically observation levels
and evaluate
the von Neumann entropy associated with reconstructed generalized canonical
density operators. If at some stage of the extension of observation levels
 the von Neumann entropy becomes zero, it then means that the given
observation is complete and the pure state of the system
is completely ``measured''. Obviously this does not mean that this
observation level is the minimal one\footnote{In fact, in the case of
pure states there always exist just one observable (at least in a sense
of the Hermitian operator) such that the given pure state is an eigenstate
of this observable. Unfortunately, it is impossible to specify this
operator prior the complete reconstruction.}.  In the case when the
measured system is prepared in an unknown  statistical mixture it
is impossible to specify the minimal observation level prior
the measurement on the complete observation level is performed. If this
is done then by  a sequence of reductions under the condition that
the von Neumann entropy is unchanged  one can specify the minimal
observation level.

In the following sections we will apply the Jaynes principle for the
reconstruction of pure spin states. Firstly, for illustrative purposes
we present the simple example of the reconstruction of states of
a single spin-1/2 system with the help of the maximum-entropy principle.
Then we will discuss the partial reconstruction of entangled spin states.
In particular, we will analyze the problem how
to identify incomplete observation levels
on which the complete reconstruction can be performed
for the Bell and the Greenberger--Horne--Zeilinger states (i.e.,
the corresponding entropy is equal to zero
and the generalized canonical density operator is identical to
$\hat\rho_0$).


\section{A SINGLE SPIN--1/2}

Firstly we illustrate the application of the maximum-entropy  principle
for the partial quantum--state reconstruction of single spin--$1/2$ system.
Let us	consider an ensemble of
spins-$1/2$ in an unknown pure state $|\psi_0\rangle$.
In the most general case this unknown state vector $|\psi_0\rangle$
can be parameterized as
\be
|\psi_0\r= \cos\theta |1\r +\mbox{e}^{i\varphi}\sin\theta |0\r
\label{2.1}
\ee
where $|0\r$, $|1\r$  are eigenstates of the $z$-component of the
spin operator $\hat{S}_z={1\over 2}\hat{\sigma}_z$
with eigenvalues $-{1\over 2}$, ${1\over 2}$, respectively.
The corresponding density operator $\hat{\rho}_0=|\psi_0\r \l\psi_0|$
can be written in the form
\be
\hat{\rho}_0= {1\over 2}\left( \hat{I} + \vec{n}.\hat{\vec{\sigma}}
\right)
\label{2.1b}
\ee
where $\hat{I}$ is the unity operator,
$\vec{n}=(\sin 2\theta \cos\varphi,\sin 2\theta \sin\varphi,\cos 2\theta)$;
$\hat{\vec{\sigma}}=(\hat{\sigma}_x,\hat{\sigma}_y,\hat{\sigma}_z)$
are the Pauli spin operators which in the matrix representation in the basis
$|0\r$, $|1\r$ read
\be
\hat{\sigma}_x=\left(\begin{array}{cc} 0 & 1 \\ 1 & 0 \end{array}\right),
\quad
\hat{\sigma}_y=\left(\begin{array}{cc} 0 & -i \\ i & 0 \end{array}\right),
\quad
\hat{\sigma}_z=\left(\begin{array}{cc} 1 & 0 \\ 0 & -1 \end{array}\right)
.\label{2.1c}
\ee

To determine completely the unknown state one has to measure
three linearly independent (e.g., orthogonal) projections of the spin.
After the measurement of the expectation value of each observable,
a reconstruction
of the generalized canonical density operator (\ref{1.4})
according to the maximum-entropy principle can be performed.
In  Table~1 we consider three observation levels defined as
${\cal O}_A^{(1)}=\{\hat\sigma_z \}$,
${\cal O}_B^{(1)}=\{ \hat\sigma_z,\hat\sigma_x \}$ and
${\cal O}_C^{(1)}=\{\hat\sigma_z,\hat\sigma_x,\hat\sigma_y \}
\equiv{\cal O}_{comp}$
[the superscript of the observation levels indicates the number of spins--$1/2$
under consideration].

Using algebraic properties of the $\hat\sigma_\nu$-operators, the
generalized canonical density operator (\ref{1.4})
 can be expressed as
\be
\hat\rho_{\cal O}={1\over Z}
\exp (-\vec{\lambda}.\hat{\vec{\sigma}})={1\over Z}
\left[ \cosh|\lambda| \hat{I} - \sinh|\lambda|
{ \vec{\lambda}.\hat{\vec{\sigma}}\over |\lambda| } \right],
\qquad Z=2 \cosh|\lambda|
\label{2.2}
\ee
with $\vec{\lambda}=(\lambda_x,\lambda_y,\lambda_x)$ and
$|\lambda|^2=\lambda_x^2+\lambda_y^2+\lambda_z^2$.
The final form of the $\hat\rho_{\cal O}$ on  particular
observation levels is given in Table~1.
The corresponding entropies can be written as
\be
S_{\cal O}=-p_{\cal O} \ln p_{\cal O} -
	   (1-p_{\cal O}) \ln (1-p_{\cal O})
\ee
where $p_{\cal O}$ is one eigenvalue of $\hat\rho_{\cal O}$
[the other eigenvalue is equal to $(1-p_{\cal O})$] which
reads as
\be
p_A={1+|\l\hat{\sigma}_z\r| \over 2}, \quad
p_B={1+\sqrt{\l\hat{\sigma}_x\r^2+\l\hat{\sigma}_z\r^2} \over 2}, \quad
p_{\rm comp}={1+\sqrt{\l\hat{\sigma}_x\r^2+\l\hat{\sigma}_y\r^2+
\l\hat{\sigma}_z\r^2} \over 2}
.\ee
It is seen that the entropy $S_{\cal O}$ is equal to zero if and
only if $p_{\cal O}=1$. From here follows that on ${\cal O}_A^{(1)}$ only
the basis vectors $|0\r$ and $|1\r$ with $|\l\hat{\sigma}_z\r|=1$
can be fully reconstructed. Nontrivial is
${\cal O}_B^{(1)}$, on which a whole set of pure states (\ref{2.1})
with $\l \hat{\sigma}_y\r=0$ (i.e., $\varphi=0$)
can be uniquely determined. For such states $S_B=0$ and further
measurement
of the $\hat\sigma_y$ on ${\cal O}_{comp}$ represents redundant
(useless) information.


\section{TWO SPINS--1/2}

Now we assume a system composed of two {\em distinguishable} spins--$1/2$.
If we are performing only {\em local} measurements
of observables such as $\hat{\sigma}_\mu^{(1)}\otimes\hat{I}^{(2)}$
and $\hat{I}^{(1)}\otimes\hat{\sigma}_\nu^{(2)}$ (here	superscripts label
the particles) which do not reflect correlations between the particles
then the reconstruction of the density operator reduces to an estimation
 of individual (uncorrelated)
spins--1/2, i.e., the reconstruction reduces to the problem discussed
in the previous section.  For each spin--$1/2$
the reconstruction can be performed separately and the resulting
generalized canonical density operator is given as a tensor product of
particular generalized canonical density operators, i.e.,
$\hat\rho=\hat\rho^{(1)}\otimes \hat\rho^{(2)}$.
In this case just the uncorrelated states
$|\psi_0\r=|\psi_0^{(1)}\r \otimes |\psi_0^{(2)}\r$ can be fully
reconstructed.
Nevertheless, the correlated (nonfactorable) states
$|\psi_0\r \neq |\psi_0^{(1)}\r \otimes |\psi_0^{(2)}\r$
are of central interest.

In general, any density operator of a system composed of two
distinguishable
spins--$1/2$ can be represented by a $4\times 4$ Hermitian matrix
and $15$ independent numbers are required for its complete
determination. It is worth noticing that $15$ operators (observables)
\be
\{ \hat{\sigma_\mu}^{(1)}\otimes\hat{I}^{(2)},
\hat{I}^{(1)}\otimes\hat{\sigma_\nu}^{(2)},
\hat{\sigma}_\mu^{(1)} \otimes\hat{\sigma}_\nu^{(2)} \}
\qquad (\mu,\nu=x,y,z)
\label{3.1}
\ee
together with the identity operator $\hat{I}^{(1)}\otimes\hat{I}^{(2)}$
form an operator algebra basis in which any operator can be expressed.
In this ``operator'' basis each density operator can be written as
\be
\hat{\rho}= {1\over 4}\left[ \hat{I}^{(1)}\otimes \hat{I}^{(2)} +
\vec{n}^{(1)}. ~\hat{\vec{\sigma}}^{(1)}\otimes\hat{I}^{(2)} +
\vec{n}^{(2)}. ~\hat{I}^{(1)}\otimes\hat{\vec{\sigma}}^{(2)} +
\sum_{\mu,\nu} \xi_{\mu\nu}
\hat{\sigma}_\mu^{(1)} \otimes\hat{\sigma}_\nu^{(2)} \right]
\label{3.2}
\ee
with $\xi_{\mu\nu}=
\l \hat{\sigma}_\mu^{(1)}\otimes\hat{\sigma}_\nu^{(2)} \r$
($\mu, \nu = x, y, z$).

Using the maximum-entropy principle we can (partially) reconstruct an unknown
density operator $\hat\rho_0$ on various observation levels.
Conceptually the method of maximum entropy is rather straightforward:
one has to express the generalized canonical density operator (\ref{1.4})
for two spins-1/2  in the form (\ref{3.2}) from which a set
of nonlinear equations for Langrange multipliers $\lambda_\nu$ is obtained.
Due to algebraic properties of the operators under the consideration the
practical realization of this programme can be technically  difficult
(see Appendix A).

In Table~2 we define some nontrivial observation
levels.
Measured observables which define a particular observation
level are indicated in Table~2 by bullets ($\bullet$)
 while the empty circles ($\circ$) indicate
unmeasured observables (i.e., these observables are not included in the
given observation level)
 for which the maximum-entropy principle ``predicts''
 nonzero mean values.
This means that the maximum-entropy principle
provide us with a nontrivial estimation of mean values of unmeasured
observables. The generalized canonical density operators which correspond
to the observation levels considered in Table~2 are presented
in Table~3. The signs ``$\oplus,\ominus$'' are used to indicate
unmeasured observables for which nontrivial information can be obtained
with the help of  the maximum-entropy principle.

\subsection{Reconstruction of Bell states}
In what follows we analyze a partial reconstruction of the Bell states
(i.e., the most correlated two particle states)
on observation levels given in Table~2. One  of our
main tasks will be
to find the minimum observation level (i.e., the set of system
observables) on which the complete reconstruction of these states
can be performed. Obviously, if all 15 observables are measured,
then any state of two spins-1/2  can be reconstructed precisely.
Nevertheless, due to the quantum entanglement between the two particles,
measurements of some observables will simply be redundant.
To find the minimal set of observables which  uniquely determine
the Bell state one has to  perform either a sequence of reductions of the
complete observation level, or	a systematic extension
of the most trivial observation level ${\cal O}^{(2)}_{A}$.

Let us consider particular examples of Bell states, of the form
\be
|\Psi_\varphi^{  (Bell)}\rangle={1\over\sqrt{2}}
\left[ |1,1\rangle + \mbox{e}^{i\varphi} |0,0\rangle \right],
\quad \hat\rho_\varphi^{(Bell)}
=|\Psi_\varphi^{ (Bell)}\rangle \langle \Psi_\varphi^{ (Bell)}|,
\label{3.3}
\ee
(other Bell states are discussed later).
These maximally correlated states have the property that the
result of a measurement
performed on one of the two spins--$1/2$ uniquely determines the
state of the
second spin.
Therefore, these states find their applications in quantum communication
systems \cite{Ekert}.
  In addition,
they are suitable for testing fundamental
principles of quantum mechanics \cite{Peres}
such as the complementarity principle or local hidden--variable theories
\cite{Greenberger}.

Let us analyze now a sequence of successive extensions of the observation
level ${\cal O}_A^{(2)}$
\be
{\cal O}_A^{(2)}\subset{\cal O}_B^{(2)}\subset{\cal O}_C^{(2)}
\subset {\cal O}_D^{(2)}
.\label{3.4}
\ee
The observation level ${\cal O}_A^{(2)}$ (see Tab.2)
is associated with the measurement
of $\hat{\sigma}_z$ observables of each spin individually, i.e.,
it is insensitive with respect to correlations between the spins.
On ${\cal O}_B^{(2)}$ both $z$-spin components of particular spins
and their correlation have been recorded (simultaneous
measurement of these observables is possible because they commute).
Further extension to the observation level
on ${\cal O}_C^{(2)}$ corresponds to a rotation of
the Stern--Gerlach apparatus such that the $x$-spin component of the
 second spin--$1/2$ is measured.
The observation level ${\cal O}_D^{(2)}$ is associated with
another rotation of the Stern--Gerlach apparatus which would allow us
to measure the $y$-spin
component.
The generalized canonical density operators on the observation levels
${\cal O}_B^{(2)}$, ${\cal O}_C^{(2)}$ and ${\cal O}_D^{(2)}$
predict zero mean values for all the unmeasured observables (\ref{3.1})
(see  Table~3).

In general,    successive extensions (\ref{3.4})
of the observation level
${\cal O}_A^{(2)}$ should be accompanied by a decrease in the
entropy of the reconstructed state which should reflect increase of our
knowledge about the quantum-mechanical system under consideration.
Nevertheless, we note that
 there are states for which the entropy remains
constant when  ${\cal O}_B^{(2)}$ is extended towards ${\cal O}_C^{(2)}$ and
${\cal O}_D^{(2)}$, i.e., the performed measurements are in fact redundant.
For instance, this is the case for the maximally correlated state (\ref{3.3}).
Here entropies associated with given observation levels read
\be
S_A=2\ln 2, \qquad S_B=S_C=S_D=\ln 2
,\ee
respectively, which
mean that these observation levels are not suitable for reconstruction
of the Bell states. The reason is that the Bell states
have no ``preferable'' direction for each individual spin,
i.e., $\langle \hat{\sigma}_{\mu}^{(p)} \rangle =0$ for  $\mu=x,y,z$
and $p=1,2$.

From the above it follows that, for a nontrivial  reconstruction
of Bell states, the observables
which reflect correlations between composite spins also have to be included
into the observation level.
Therefore let us now discuss the sequence of observation levels
\be
  {\cal O}_E^{(2)}\subset{\cal O}_F^{(2)}\subset{\cal O}_G^{(2)}
\label{3.5}
\ee
associated with simultaneous measurement of spin components of the two
particles [see Table 2].
 The corresponding generalized canonical
density operators are given  in Table~3.
To answer the question of  which states can be completely reconstructed
on the observation level
${\cal O}_E^{(2)}$ we evaluate
the von Neumann entropy (\ref{1.3}) of the generalized canonical
density operator $\hat{\rho}_E$. For the Bell states we find that
$S_E=-p_E \ln p_E - (1-p_E) \ln (1-p_E)$ where $p_E=(1-\cos\varphi)/2$.
We can also compare directly $\hat{\rho}_\varphi^{(Bell)}$ with
$\hat{\rho}_E$. The density  operator $\hat{\rho}_\varphi^{(Bell)}$
in the matrix form can be written as
\be
\hat{\rho}_\varphi^{(Bell)}={1\over 2}\left(
\begin{array}{cccc}
1 & 0 & 0 & \mbox{e}^{-i\varphi} \\
0 & 0 & 0 & 0 \\
0 & 0 & 0 & 0 \\
\mbox{e}^{i\varphi} & 0 & 0 & 1
\end{array} \right)
,\ee
while the corresponding operator reconstructed on
the observation level ${\cal O}_E^{(2)}$ reads
\be
\hat{\rho}_E={1\over 2}\left(
\begin{array}{cccc}
1 & 0 & 0 & \cos\varphi \\
0 & 0 & 0 & 0 \\
0 & 0 & 0 & 0 \\
\cos\varphi & 0 & 0 & 1
\end{array} \right).
\ee
We see that $\hat{\rho}_\varphi^{(Bell)}=\hat{\rho}_E$
and $S[\hat{\rho}_E]=0$ only
if $\varphi=0$ or $\pi$ which means that
the Bell states $|\Psi_{\varphi=0,\pi}\rangle={1\over\sqrt{2}}
\left[ |1,1\rangle \pm |0,0\rangle \right]$ are completely determined
by mean values of two observables
$\hat{\sigma}_z^{(1)}\otimes \hat{\sigma}_z^{(2)}$ and
$\hat{\sigma}_x^{(1)}\otimes \hat{\sigma}_x^{(2)}$
 and that these states can be completely reconstructed on  ${\cal O}_E^{(2)}$.
We note  that two other maximally
correlated states $|\Phi_{\pm}\rangle={1\over\sqrt{2}}
\left[ |0,1\rangle \pm |1,0\rangle \right]$
can also be completely reconstructed on ${\cal O}_E^{(2)}$.

The  extension of ${\cal O}_E^{(2)}$ to ${\cal O}_F^{(2)}$ does not increase
the amount of  information about  the Bell states (\ref{3.3}) with
$\varphi\neq 0,\pi$. For this reason we have to consider
further extension of ${\cal O}_F^{(2)}$
 to the observation level ${\cal O}_G^{(2)}$
(see Table 2 and Appendix A). In what follows we will show that
this is an observation level
on which {\em all}  Bell states
(\ref{3.3}) can be completely reconstructed.
To see this one has to realize two facts. Firstly,
the generalized canonical density operator $\hat{\rho}_G$ given by
Eq.(\ref{1.4}) can be expressed as a linear superposition of observables
associated with the given observation level, i.e.:
\be
\hat{\rho}_G=\frac{1}{Z_G}\exp\left(
-\sum_{\mu=x,y,z}\lambda_{\mu\mu}\hat{\sigma}_{\mu}^{(1)}\otimes
\hat{\sigma}_{\mu}^{(2)}
-\lambda_{xy}\hat{\sigma}_x^{(1)}\otimes \hat{\sigma}_y^{(2)}
-\lambda_{yx}\hat{\sigma}_y^{(1)}\otimes \hat{\sigma}_x^{(2)}
\right)
\nonumber
\\
=\frac{1}{4}\left(\hat{1}
-\sum_{\mu=x,y,z}\xi_{\mu\mu}\hat{\sigma}_{\mu}^{(1)}\otimes
\hat{\sigma}_{\mu}^{(2)}
-\xi_{xy}\hat{\sigma}_x^{(1)}\otimes \hat{\sigma}_y^{(2)}
-\xi_{yx}\hat{\sigma}_y^{(1)}\otimes \hat{\sigma}_x^{(2)}
\right)
\ee
where the parameters $\xi_{\mu\nu}$ are functions of
the Lagrange multipliers $\lambda_{\mu\nu}$.
Secondly,
for  Bell states (\ref{3.3}) the
only observables which have nonzero expectation values are those
associated with ${\cal O}_G^{(2)}$.
 Namely,
$\l \hat{\sigma}_z^{(1)} \otimes\hat{\sigma}_z^{(2)} \r=1$,
$\l \hat{\sigma}_x^{(1)} \otimes\hat{\sigma}_x^{(2)} \r=
-\l \hat{\sigma}_y^{(1)} \otimes\hat{\sigma}_y^{(2)} \r=\cos\varphi$
and $\l \hat{\sigma}_x^{(1)} \otimes\hat{\sigma}_y^{(2)} \r=
\l \hat{\sigma}_y^{(1)} \otimes\hat{\sigma}_x^{(2)} \r=\sin\varphi$.
It means that all coefficients in  the generalized
canonical density operator $\hat\rho_G$ given by Eq. (\ref{3.2})
are uniquely determined by the
measurement, i.e., $\hat\rho_G=\hat\rho_\varphi$.

From the above it follows that	Bell states can be completely
reconstructed on the observation level ${\cal O}_G^{(2)}$.
On the other hand, ${\cal O}_G^{(2)}$ is not the {\em minimum} observation
level on which these states can  be completely reconstructed.
The minimum set of observables which would allow us to
reconstruct  Bell states uniquely can be found
by a {\em reduction} of ${\cal O}_G^{(2)}$.
Direct inspection of a	finite number of possible reductions
reveals that  Bell states can be completely reconstructed
on those observation level which can be obtained from ${\cal O}_G^{(2)}$
when one of the observables
$\hat{\sigma}_\nu^{(1)} \otimes\hat{\sigma}_\nu^{(2)}$ ($\nu=x,y,z$)
is omitted.
As an example, let us consider the observation level
${\cal O}_H^{(2)}$ given in Table~2 which
represents a reduction of ${\cal O}_G^{(2)}$ when  the observable
$\hat{\sigma}_z^{(1)} \otimes\hat{\sigma}_z^{(2)}$ is omitted.
Performing the Taylor series expansion of the generalized canonical
density operator $\hat\rho_H$ defined by Eq. (\ref{1.4}) one can find
that the only  new observable
$\hat{\sigma}_z^{(1)} \otimes\hat{\sigma}_z^{(2)}$
 enters the expression for the $\hat\rho_H$
as indicated in Table~3. The coefficient $t$ in front of
$\hat{\sigma}_z^{(1)} \otimes\hat{\sigma}_z^{(2)}$
can either be found explicitly in a closed analytical form
(see Appendix A) or can be obtained from the following variational problem.
Namely, we remind ourselves that
the expression (\ref{1.4}) for $\hat\rho_H$ helps us to identify
those unmeasured observables for which
the Jaynes principle of the maximum entropy ``predicts'' nonzero
mean values. At this stage we
still have  to find the particular value of the parameter $t$
for which the density operator $\hat\rho_H$ in Table~3 leads to the
maximum of the von Neumann entropy. To do so
we search through the one-dimensional parametric space which is bounded as
$-1 \le t\le 1$. To be specific,
first of all, for $t\in \l -1,1 \r$  we have to exclude those operators
 which are not true density operators (i.e., any such operators which have
negative eigenvalues). Then we ``pick'' up from a physical parametric
subspace the generalized canonical density operator with the maximum von
Neumann entropy. Direct calculation for  Bell states shows that the
physical parametric subspace is reduced to an isolated ``point''
with $t=\l \hat{\sigma}_z^{(1)} \otimes\hat{\sigma}_z^{(2)} \r=1$.
Therefore we conclude that
  Bell states can completely be reconstructed on ${\cal O}_H$.
Two other minimum observation levels suitable for the complete
reconstruction of  Bell states
can be obtained by a reduction
of ${\cal O}_G^{(2)}$ when either
$\hat{\sigma}_x^{(1)} \otimes\hat{\sigma}_x^{(2)}$
or $\hat{\sigma}_y^{(1)} \otimes\hat{\sigma}_y^{(2)}$ is omitted.
On the other hand, direct inspection shows that
a reduction of ${\cal O}_G^{(2)}$ by exclusion of either
$\hat{\sigma}_x^{(1)} \otimes\hat{\sigma}_y^{(2)}$
or $\hat{\sigma}_y^{(1)} \otimes\hat{\sigma}_z^{(2)}$ leads to
an incomplete observation level with respect to  Bell states.

In what follows we  discuss briefly
two other  observation levels
${\cal O}_I^{(2)}$ and ${\cal O}_J^{(2)}$ which are defined in Table~2.
The observation level ${\cal O}_I^{(2)}$ serves as an example when one
can find an analytical expression for the Taylor series expansion
of the canonical density operator $\hat\rho_I$ (\ref{1.4}) in
the form (\ref{3.2}). The coefficients (functions of the original Lagrange
multipliers) in front of particular observables in Eq.(\ref{3.2})
can be identified and are given in Table~3. Problems do appear
when  ${\cal O}_I^{(2)}$ is extended towards ${\cal O}_J^{(2)}$.
In this case we cannot simplify the exponential expression (\ref{1.4})
for $\hat\rho_J$ and rewrite it analytically in the
form (\ref{3.2}) as a linear combination of the observables (\ref{3.1}).
In this situation one should apply the following procedure:
firstly, by performing the Taylor-series expansion of the $\hat\rho_J$
to the lowest orders one can identify
the observables with nonzero coefficients
in the form (\ref{3.2}). Namely, for $\hat\rho_J$
the additional observables
$\hat{\sigma}_z^{(1)}\otimes\hat{\sigma}_x^{(2)}$,
$\hat{\sigma}_x^{(1)}\otimes\hat{\sigma}_z^{(2)}$ and
$\hat{\sigma}_y^{(1)}\otimes\hat{\sigma}_y^{(2)}$
appear in addition to those which form ${\cal O}_H^{(2)}$ [see Table~3].
The corresponding coefficients $u,v,w\in \l -1,1 \r$ form a bounded
three--dimensional parametric space $(u,v,w)$.
In the second step one can use constructively the maximum-entropy principle
to choose within this parametric space	the density operator
with the maximum von Neumann entropy. The basic procedure is to scan
the whole three--dimensional parametric space. At the beginning,
one has to select out those density operators (i.e., those
parameters $u,v,w$) which posses negative eigenvalues and do not
represent genuine density operators. Finally,
from a remaining set of  ``physical'' density operators
which are semi--positively defined  the canonical density
operator $\hat\rho_J$ with maximum von Neumann entropy has to be chosen.
For a completeness, let us notice that for  Bell states
the observation levels ${\cal O}_I^{(2)}$ and ${\cal O}_J^{(2)}$
are equivalent to ${\cal O}_E^{(2)}$, i.e.,
$\hat\rho_I=\hat\rho_J=\hat\rho_E$.

In this section we have found  the minimum observation
levels [e.g., ${\cal O}_H^{(2)}$] which are suitable for
the  complete reconstruction  of  Bell states.
These observation levels are associated
with the measurement of two--spin
correlations
$\hat{\sigma}_x^{(1)}\otimes\hat{\sigma}_z^{(2)}$,
$\hat{\sigma}_y^{(1)}\otimes\hat{\sigma}_z^{(2)}$ and
two of the observables
$\hat{\sigma}_\nu^{(1)} \otimes\hat{\sigma}_\nu^{(2)}$
($\nu=x,y,z$).
Once this problem has been solved, it is
interesting then to find a minimum set of observables suitable for
a complete reconstruction of maximally correlated spin states systems
consisting of more than two spins--$1/2$.
In the following section we will investigate the (partial) reconstruction
of Greenberger-Horne-Zeilinger states of three spins--1/2
on various observation levels.


\section{THREE SPINS--{1/2}}

Even though the Jaynes principle of  maximum entropy provides us
with general instructions on how to reconstruct density operators of
quantum-mechanical systems practical applications of this reconstruction
scheme may face serious difficulties. In many cases the reconstruction
scheme fails due to insurmountable technical problems (e.g. the system
of equations for Lagrange multipliers cannot be solved explicitly).
We have  illustrated these problems in the previous section
when we have discussed the reconstruction of a density operator of
two spins--1/2. Obviously, the general problem of reconstruction
of density operators describing a system composed of three spins--1/2
is much more difficult. Nevertheless a (partial) reconstruction of some
states of this system can be performed. In particular, in this section
we will discuss a reconstruction of the maximally correlated three
spin-1/2 states -- the so-called
{\em Greenberger-Horne-Zeilinger} (GHZ) state \cite{Greenberger}:
\be
|\Psi_\varphi^{(GHZ)}\r={1\over\sqrt{2}}
\left[ |1,1,1\r +\mbox{e}^{\mbox{i}\varphi} |0,0,0\r \right], \quad
\hat\rho_\varphi^{(GHZ)}=|\Psi_\varphi \r \l \Psi_\varphi |
.\label{4.1}
\ee
Our main task will be to identify, with the help of the Jaynes principle of
 maximum entropy,  the minimum observation
level on which the GHZ state can be completely reconstructed.

We start with a relatively simple observation level ${\cal O}_B^{(3)}$
such that only
{\em two}-particle correlations of the neighboring spins are measured,
i.e.
\be
{\cal O}_B^{(3)}=\{
\hat{\sigma}_z^{(1)}\otimes \hat{\sigma}_z^{(2)}\otimes \hat{I}^{(3)},
\hat{I}^{(1)}\otimes\hat{\sigma}_z^{(2)}\otimes \hat{\sigma}_z^{(3)}
\}.
\ee
The generalized density operator associated with this observation level
reads
\be
\hat{\rho}_B= &{1\over 8}& \left[
\hat{I}^{(1)}\otimes \hat{I}^{(2)}\otimes \hat{I}^{(3)}
+ \l \hat{\sigma}_z^{(1)}\otimes\hat{\sigma}_z^{(2)}\otimes \hat{I}^{(3)} \r
\hat{\sigma}_z^{(1)} \otimes \hat{\sigma}_z^{(2)} \otimes \hat{I}^{(3)}
\right. \nonumber \\
&+&
\l \hat{I}^{(1)}\otimes \hat{\sigma}_z^{(2)}\otimes\hat{\sigma}_z^{(3)}\r
\hat{I}^{(1)} \otimes \hat{\sigma}_z^{(2)}\otimes \hat{\sigma}_z^{(3)}
\nonumber \\
&\oplus& \left.
\l \hat{\sigma}_z^{(1)}\otimes\hat{\sigma}_z^{(2)}\otimes \hat{I}^{(3)} \r
\l \hat{I}^{(1)}\otimes \hat{\sigma}_z^{(2)}\otimes\hat{\sigma}_z^{(3)} \r
\hat{\sigma}_z^{(1)}\otimes \hat{I}_{2} \otimes \hat{\sigma}_z^{(3)}
\right]
.\label{4.2}
\ee
where `$\oplus$'' indicates a prediction for the unmeasured observable.
In particular, for the GHZ states (\ref{4.1}) we obtain the following
generalized canonical density operator
\be
\hat{\rho}_B^{(GHZ)}&=& {1\over 8}\left[
\hat{I}^{(1)}\otimes \hat{I}^{(2)}\otimes \hat{I}^{(3)}
\right. \nonumber \\
&+& \left.
\hat{\sigma}_z^{(1)} \otimes \hat{\sigma}_z^{(2)} \otimes \hat{I}^{(3)}
+ \hat{I}^{(1)}\otimes \hat{\sigma}_z^{(2)} \otimes \hat{\sigma}_z^{(3)}
\oplus \hat{\sigma}_z^{(1)}\otimes \hat{I}_{2} \otimes \hat{\sigma}_z^{(3)}
\right] \nonumber \\
&=& {1\over 2} |1,1,1\r \l 1,1,1| + {1\over 2} |0,0,0\r \l 0,0,0|
.\label{4.3}
\ee
The
reconstructed density operator $\hat{\rho}_B^{(GHZ)}$ describes a mixture
of three-particle states and it
does not
contain any information about the three-particle correlations associated
with the GHZ states. In other words, on ${\cal O}_B^{(3)}$
 the phase information
which plays essential role for a description of quantum entanglement
cannot be reconstructed. This is due to the fact that
 the density
operator $\hat{\rho}_B^{(GHZ)}$  is equal to the phase-averaged
GHZ density operator, i.e.
\be
\hat\rho_B^{(GHZ)}=\frac{1}{2\pi}\int_{-\pi}^{\pi}
\hat\rho_\varphi^{(GHZ)} \, d\varphi.
\ee
Because of this loss of information,
the von\,Neumann entropy of  the state $\hat\rho_B^{(GHZ)}$
is equal to $\ln 2$. We note, that when the GHZ states
are reconstructed on the observation levels
$ {\cal O}_{B'}^{(3)}=\{
\hat{\sigma}_\mu^{(1)}\otimes \hat{\sigma}_\mu^{(2)}\otimes \hat{I}^{(3)},
\hat{I}^{(1)}\otimes\hat{\sigma}_\mu^{(2)}\otimes \hat{\sigma}_\mu^{(3)}
\}$ ($\mu=x,y$),
then the corresponding reconstructed operators are again given by
Eq.(\ref{4.3}). These examples illustrate the fact that three-particle
correlation cannot be in general reconstructed via the measurement
of two-particle correlations.

To find the observation level on which the complete reconstruction
of the GHZ states can be performed we recall
the observables which may have nonzero mean values for these states.
Using abbreviations
\be
&\xi_{\mu_1 \nu_2}=\l \hat\sigma_\mu^{(1)} \otimes \hat\sigma_\nu^{(2)}
\otimes \hat{I}^{(3)} \r, \quad
\xi_{\mu_2 \nu_3}=\l \hat{I}^{(1)}\otimes \hat\sigma_\mu^{(2)}
\otimes \hat\sigma_\nu^{(3)} \r, \quad
\xi_{\mu_1 \nu_3}=\l \hat\sigma_\mu^{(1)} \otimes\hat{I}^{(2)}
\otimes \hat\sigma_\nu^{(3)} \r,& \nonumber \\
&\zeta_{\mu_1 \nu_2 \omega_3} = \l  \hat\sigma_\mu^{(1)}
\otimes \hat\sigma_\nu^{(2)} \otimes \hat\sigma_\omega^{(3)} \r,
\qquad (\mu,\nu,\omega=x,y,z), &
\ee
we find the nonzero mean values to be
\be
\xi_{z_1 z_2} &=&\xi_{z_2 z_3}=\xi_{z_1 z_3}=1, \nonumber \\
\zeta_{x_1 x_2 y_3} &=& \zeta_{y_1 x_2 x_3}=
\zeta_{x_1 y_2 x_3}= \sin\varphi,   \nonumber \\
\zeta_{y_1 y_2 x_3} &=& \zeta_{x_1 y_2 y_3}=
\zeta_{y_1 x_2 y_3}= -\cos\varphi,  \nonumber \\
\zeta_{x_1 x_2 x_3} &=& \cos\varphi, \nonumber \\
\zeta_{y_1 y_2 y_3} &=&-\sin\varphi
\label{4.4a}.\ee
We  see that for arbitrary $\varphi$ there exist non--vanishing
three--particle correlations $\zeta_{\mu_1 \nu_2 \omega_3}$.
The observation level which consists
of all the observables with nonzero mean values is the complete
observation level with respect to the GHZ states.
Our task now is to reduce this set of observables to a minimum
observation level on which the GHZ states can  still be uniquely determined.
In practice it means that each observation level which is suitable
for the detection of  the existing coherence and correlations
should incorporate some of the observables with nonzero mean
values. The other observables of these observation levels
should	result as a consequence of mutual tensor products
which appear in the Taylor series expansion of the generalized canonical
density operator (\ref{1.4}). It can be seen
 by direct inspection of  the finite number of possible reductions
that the minimum set of the observables which matches
these requirements consists of two two--spin observables and
two three--spin observables. For the illustration we consider the
 observation level
\be
{\cal O}_C^{(3)}=\{
\hat{\sigma}_z^{(1)}\otimes\hat{\sigma}_z^{(2)}\otimes\hat{I}^{(3)},
\hat{I}^{(1)}\otimes\hat{\sigma}_z^{(2)}\otimes\hat{\sigma}_z^{(3)},
\hat{\sigma}_x^{(1)}\otimes\hat{\sigma}_x^{(2)}\otimes\hat{\sigma}_x^{(3)},
\hat{\sigma}_y^{(1)}\otimes\hat{\sigma}_y^{(2)}\otimes\hat{\sigma}_y^{(3)}
\}. \label{4.4}
\ee
In this case
the exponent $\hat{C}$ of the generalized canonical density operator
$\hat\rho_C=\exp(-\hat{C})/Z_C$ [see Eq.(\ref{1.4})] can be rewritten as
$\hat{C}=\hat{C}_1+\hat{C}_2$ with $\hat{C}_1=\gamma_{12}
\hat{\sigma}_z^{(1)}\otimes\hat{\sigma}_z^{(2)}\otimes\hat{I}^{(3)}
+\gamma_{23}
\hat{I}^{(1)}\otimes\hat{\sigma}_z^{(2)}\otimes\hat{\sigma}_z^{(3)}$
and $\hat{C}_2=\alpha
\hat{\sigma}_x^{(1)}\otimes\hat{\sigma}_x^{(2)}\otimes\hat{\sigma}_x^{(3)}
+\beta
\hat{\sigma}_y^{(1)}\otimes\hat{\sigma}_y^{(2)}\otimes\hat{\sigma}_y^{(3)}$.
The operators $\hat{C}_1$, $\hat{C}_2$ commute and further calculations
are straightforward. After some algebra the generalized density operator
$\hat\rho_C$ can be found in the form
\be
&&\hat{\rho}_C=
{1\over 8} \left[
\hat{I}^{(1)}\otimes \hat{I}^{(2)}\otimes \hat{I}^{(3)} +
\xi_{z_1 z_2} \hat{\sigma}_z^{(1)}\otimes\hat{\sigma}_z^{(2)}
\otimes\hat{I}^{(3)} +
\xi_{z_2 z_3} \hat{I}^{(1)}\otimes\hat{\sigma}_z^{(2)}
\otimes\hat{\sigma}_z^{(3)}   \right.
\label{4.5} \\	 
&& + \zeta_{x_1 x_2 x_3} \hat{\sigma}_x^{(1)}\otimes\hat{\sigma}_x^{(2)}
\otimes\hat{\sigma}_x^{(3)} +
\zeta_{y_1 y_2 y_3} \hat{\sigma}_y^{(1)}\otimes\hat{\sigma}_y^{(2)}
\otimes\hat{\sigma}_y^{(3)}
\oplus \xi_{z_1 z_2}  \xi_{z_2 z_3}
\hat{\sigma}_z^{(1)}\otimes\hat{I}^{(2)}\otimes\hat{\sigma}_z^{(3)}
\nonumber \\
&& \ominus
\zeta_{x_1 x_2 x_3} \left(
\xi_{z_1 z_2} \hat{\sigma}_y^{(1)}\otimes\hat{\sigma}_y^{(2)}\otimes
\hat{\sigma}_x^{(3)} +
\xi_{z_2 z_3} \hat{\sigma}_x^{(1)}\otimes\hat{\sigma}_y^{(2)}\otimes
\hat{\sigma}_y^{(3)} +
\xi_{z_1 z_2} \xi_{z_2 z_3}
\hat{\sigma}_y^{(1)}\otimes\hat{\sigma}_x^{(2)}\otimes
\hat{\sigma}_y^{(3)}
\right) \nonumber\\
&& \ominus \left.
\zeta_{y_1 y_2 y_3} \left(
\xi_{z_1 z_2} \hat{\sigma}_x^{(1)}\otimes\hat{\sigma}_x^{(2)}\otimes
\hat{\sigma}_y^{(3)} +
\xi_{z_2 z_3} \hat{\sigma}_y^{(1)}\otimes\hat{\sigma}_x^{(2)}\otimes
\hat{\sigma}_x^{(3)} +
\xi_{z_1 z_2} \xi_{z_2 z_3}
\hat{\sigma}_x^{(1)}\otimes\hat{\sigma}_y^{(2)}\otimes
\hat{\sigma}_x^{(3)}
\right) \right] \nonumber
.\ee
For the GHZ states
the von\,Neumann entropy of the generalized canonical density operator
$\hat{\rho}_C$ is equal to zero, from which it follows that
$\hat{\rho}_C=\hat{\rho}^{(GHZ)}_\varphi$ [see Eq.(\ref{4.1})], i.e.,
the GHZ states can be completely reconstructed on ${\cal O}_C^{(3)}$.
Moreover, the  observation level ${\cal O}_C$ represents
the {\em minimum} set of observables for complete determination
of the GHZ states.


\section{CONCLUSIONS}

We have investigated the problem of a (partial) reconstruction of
correlated spin states on different observation levels. We have found
the minimal set of observables for the complete reconstruction of
the most correlated states for systems composed of two and three
spins--$1/2$, i.e.,  Bell states and GHZ states.
Direct generalization to systems of more spins--$1/2$ is
possible.

The concept of observation levels and the maximum-entropy principle
is a powerful tool which can be used also for other physical
system, e.g.,  for the reconstruction of the states of a monochromatic
light--field \cite{Buzek}.
We recall that this reconstruction scheme is based on the knowledge
of the {\em exact} mean values of given observables
or their probability  distributions (see Appendix A).
Theoretically, this means that an {\em infinite}
number of measurement over an ensemble of identically prepared system
has to be performed in order to obtain those mean values which are needed.
In practice, if the number of measurements is sufficiently high, then
the mean values can be considered to be measured precisely enough
and the Jaynes principle can be applied for a state reconstruction.
On the other hand, if just  few measurements are performed, then
the mean values of the considered observables are not known
and the Jaynes principle cannot be used.

In this case another reconstruction scheme has to be applied. In particular,
in the case of a small number of measurements  the
Bayesian reconstruction scheme \cite{Jones} can be effectively utilized.
We will address the problem of a reconstruction of correlated spin
systems based on Bayesian methods elsewhere  \cite{Derka}.

\vspace{1.5truecm}

{\bf Acknowledgements}\newline
This work was in part supported by the United Kingdom Engineering
and Physical Sciences Research Council, the Max-Planck Society
and the Grant Agency
VEGA of the Slovak Academy of Sciences.
We acknowledge the support by the  East-West Program of the Austrian
Academy of Sciences under the contract No. 45.367/6-IV/3a/95 of the
\"{O}sterreichisches Bundesministerium f\"{u}r Wissenschaft und Forschung.


\section*{Appendix A}
Conceptually the reconstruction scheme based on the Jaynes principle
of the maximum entropy is very simple. On the other hand particular
analytical calculations can be difficult and in many cases cannot
be performed. In this appendix we present explicit calculations
of generalized canonical density operators (GCDO) and corresponding
entropies for two observation levels
${\cal O}_G^{(2)}$ and ${\cal O}_H^{(2)}$ defined in  Table 2.

\subsection*{ A. 1. Observation level ${\cal O}_G^{(2)}$}
Let as assume the observation level  ${\cal O}_G^{(2)}$
given by the set of observables
$\{\hat \sigma _z^{(1)}\otimes\hat\sigma _z^{(2)};
\hat\sigma _x^{(1)}\otimes\hat\sigma _x^{(2)};
\hat \sigma _x^{(1)}\otimes\hat \sigma _y^{(2)};
\hat\sigma _y^{(1)}\otimes\hat\sigma _x^{(2)};
\hat \sigma _y^{(1)}\otimes\hat \sigma _y^{(2)}\}$. In this case
the GCDO reads
$$
\hat\rho_{ G}={1 \over {Z_G}}\exp \left( {-\hat E} \right)
\eqno({\rm A.1})
$$
where
$$
Z_G= \mbox{Tr}\left[ {\exp \left( {-\hat E} \right)} \right]
\eqno({\rm A.2}) $$
is the partition function. Here we have used the abbreviation
$$
\hat E=\lambda _{zz}\hat \sigma _z^{(1)}\otimes\hat\sigma _z^{(2)}
+\lambda _{xx}
\hat\sigma _x^{(1)}\otimes\hat\sigma _x^{(2)}+\lambda _{xy}
\hat \sigma _x^{(1)}\otimes\hat \sigma _y^{(2)}
+\lambda _{yx}\hat\sigma _y^{(1)}\otimes\hat\sigma _x^{(2)}
+\lambda _{yy}\hat \sigma _y^{(1)}\otimes\hat \sigma _y^{(2)}
.\eqno({\rm A.3}) $$
The corresponding entropy has the form
$$
S_G=\ln Z_G+\lambda _{zz}\xi _{zz}+\lambda _{xx}\xi _{xx}
+\lambda _{xy}\xi _{xy}
+\lambda _{yx}\xi _{yx}+\lambda _{yy}\xi _{yy}
,\eqno({\rm A.4})
$$
Using the algebraic properties of the operators associated with the given
observation level we find  the GCDO (A.1) to read
$$
\hat\rho_{ G}={1 \over 4}\left[ \hat I^{(1)}\otimes\hat I^{(2)}+
\xi _{zz}\hat\sigma _z^{(1)}\otimes \hat\sigma _z^{(2)}
+\xi _{xx}\hat \sigma _x^{(1)} \otimes\hat \sigma _x^{(2)} \right.
$$ $$
\left. +\xi _{xy}\hat \sigma _x^{(1)} \otimes\hat \sigma _y^{(2)}
+\; \xi _{yx}\hat \sigma _y^{(1)}\otimes \hat\sigma _x^{(2)}
+\xi _{yy} \hat\sigma _y^{(1)}\otimes \hat\sigma _y^{(2)}\right]
\eqno({\rm A.5})
$$
where we use the notation
$$
\xi _{\mu\nu}\equiv \left\langle {\hat \sigma _\mu^{(1)}
\otimes\hat \sigma _\nu^{(2)}} \right\rangle, \qquad (\mu,\nu=x,y,z).
\eqno({\rm A.6}) $$

Now we	express the entropy as a function of expectation
values of operators associated with the observation level ${\cal O}_G^{(2)}$.
With the help of this entropy function we can perform reductions of
${\cal O}_G^{(2)}$ to the observation levels
${\cal O}_H^{(2)}$, ${\cal O}_F^{(2)}$ and ${\cal O}_E^{(2)}$.
In order to perform this reduction we
express $\lambda _{\mu\nu}$ in Eq. ({\rm A.4}) as functions of
the expectation values $\xi _{\mu\nu}$. To do so we utilize the relation
$$
\xi _{\mu\nu}=-{{\partial \ln Z_G} \over {\partial \lambda _{\mu\nu}}}
.\eqno({\rm A.7})
$$
The partition function $Z_G$ can be found when we rewrite the operator
$\hat{E}$ in Eq.(A.4) as a 4$\times$4 matrix:
$$
\hat E=\left( {\matrix{a&0&0&{d^*}\cr
0&{-a}&{b^*}&0\cr
0&b&{-a}&0\cr
d&0&0&a\cr
}} \right)
,\eqno({\rm A.8})
$$
where we used the abbreviations
$$
  a=\lambda _{zz} , \quad
  b=\lambda _{xx}+\lambda _{yy}-i(\lambda _{xy}-\lambda _{yx}), \quad
  d=\lambda _{xx}-\lambda _{yy}+i(\lambda _{xy}+\lambda _{yx})
.\eqno({\rm A.9})
$$
The powers of the operator $\hat{E}$ can be written as
$$
\hat E^n=\left( {\matrix{{E_{11}^{(n)}}&0&0&{E_{14}^{(n)}}\cr
0&{E_{22}^{(n)}}&{E_{23}^{(n)}}&0\cr
0&{E_{32}^{(n)}}&{E_{33}^{(n)}}&0\cr
{E_{41}^{(n)}}&0&0&{E_{44}^{(n)}}\cr
}} \right)
,\eqno({\rm A.10})
$$
with the matrix elements given by the relations
$$
\begin{array}{rcl}
E_{11}^{(n)} &= & E_{44}^{(n)}=
{1 \over 2}\left[ {\left( {a+\left| d \right|} \right)^n
+\left( {a-\left| d \right|} \right)^n} \right] \, ,\\
E_{14}^{(n)} & = & {1 \over 2}\left[ {\left( {a+\left| d \right|} \right)^n
-\left( {a-\left| d \right|} \right)^n} \right]{{d^*}
\over {\left| d \right|}} \, , \\
E_{22}^{(n)} & = & E_{33}^{(n)}=
{1 \over 2}\left[ {\left( {-a+\left| b \right|} \right)^n
+\left( {-a-\left| b \right|} \right)^n} \right] \, ,\\
E_{23}^{(n)} & = & {1 \over 2}\left[ {\left( {-a+\left| b \right|} \right)^n
-\left( {-a-\left| b \right|} \right)^n} \right]{{b^*}
\over {\left| b \right|}} \, ,\\
   E_{32}^{(n)} & = & E_{23}^{(n)*} \, ,\\
   E_{41}^{(n)}& = & E_{14}^{(n)*} \, .
\end{array}
\eqno({\rm A.11})
$$
Now we find
$$
\exp \left( {-\hat E} \right)=\left( {\matrix{{e^{-a}\cosh\left| d
\right|}&0&0&
{-e^{-a}\sinh(\left| d \right|){{d^*} \over {\left| d \right|}}}\cr
0&{e^a\cosh\left| b \right|}&{-e^a\sinh(\left| b \right|){{b^*}
\over {\left| b \right|}}}&0\cr
0&{-e^a\sinh(\left| b \right|){{b^{}} \over {\left| b \right|}}}&
{e^a\cosh\left| b \right|}&0\cr
{-e^{-a}\sinh(\left| d \right|){{d^{}} \over {\left| d \right|}}}&
0&0&{e^{-a}\cosh\left| d \right|}\cr
}} \right)
\eqno({\rm A.12})
$$
from which we obtain the expression for  the partition function $Z_G$
$$
Z_G=2e^{-a}\cosh\left| d \right|+2e^a\cosh\left| b \right| .
\eqno({\rm A.13})
$$
For the  expectation values given by Eq.(A.7) we obtain
$$
\begin{array}{rcl}
\xi _{zz} & = & {1 \over {Z_G}}\left[ {2e^{-a}\cosh\left| d \right|
-2e^a\cosh\left| b \right|} \right] \, ;\\
\xi _{xx}& =& -{1 \over {Z_G}}\left[ {2e^{-a}\sinh(\left| d \right|){{1^{}}
\over
{\left| d \right|}}\left( {\lambda _{xx}-\lambda _{yy}}
\right)+2e^a\sinh(\left| b \right|){{1^{}} \over {\left| b \right|}}
\left(
{\lambda _{xx}+\lambda _{yy}} \right)} \right] \, ; \\
\xi _{xy} & = &-{1 \over {Z_G}}\left[ {2e^{-a}\sinh(\left| d \right|)
{{1^{}}
  \over {\left| d \right|}}\left( {\lambda _{xy}+\lambda _{yx}}
\right)
  +2e^a\sinh(\left| b \right|){{1^{}} \over {\left| b \right|}}\left(
  {\lambda _{xy}-\lambda _{yx}} \right)} \right] \, ; \\
\xi _{yx} & = &-{1 \over {Z_G}}\left[ {2e^{-a}\sinh(\left| d \right|)
{{1^{}}
  \over {\left| d \right|}}\left( {\lambda _{xy}+\lambda _{yx}} \right)
  -2e^a\sinh(\left| b \right|){{1^{}} \over {\left| b \right|}}\left(
{\lambda _{xy}
  -\lambda _{yx}} \right)} \right] \, ;\\
\xi _{yy} & = & -{1 \over {Z_G}}\left[ {-2e^{-a}\sinh(\left| d \right|)
{{1^{}}
  \over {\left| d \right|}}\left( {\lambda _{xx}-\lambda _{yy}} \right)
  +2e^a\sinh(\left| b \right|){{1^{}} \over {\left| b \right|}}\left(
{\lambda _{xx}
  +\lambda _{yy}} \right)} \right] \,.
\end{array}
\eqno({\rm A.14})
$$
If we introduce the abbreviations
$$
B=\xi _{xx}+\xi _{yy}-i\left( {\xi _{xy}-\xi _{yx}} \right) \, , \quad
D=\xi _{xx}-\xi _{yy}+i\left( {\xi _{xy}+\xi _{yx}} \right)
\eqno({\rm A.15})
$$
then with the help of Eq. (A.14) we obtain
$$
B=-{4 \over {Z_G}}e^a\sinh(\left| b \right|){{b^{}} \over {\left| b
\right|}}
\, , \quad
D=-{4 \over {Z_G}}e^{-a}\sinh(\left| d \right|){{d^{}} \over
{\left| d \right|}}
.\eqno({\rm A.16})
$$
Taking into account that
$$
\left| B \right|={4 \over {Z_G}}e^a\sinh(\left| b \right|) \, , \quad
\left| D \right|={4 \over {Z_G}}e^{-a}\sinh(\left| d \right|)
\eqno({\rm A.17})
$$
we find
$$
{B \over {\left| B \right|}}=-{b \over {\left| b \right|}}
 \, , \quad
{D \over {\left| D \right|}}=-{d \over {\left| d \right|}}
.\eqno({\rm A.18})
$$
Now we introduce four new parameters $M_i$
$$
M_1=1+\xi _{zz}+\left| D \right|  \, , \quad
  M_2=1+\xi _{zz}-\left| D \right|  \, , \\
$$
$$
  M_3=1-\xi _{zz}+\left| B \right|  \, , \quad
  M_4=1-\xi _{zz}-\left| B \right|
\eqno({\rm A.19})
$$
in terms of which we can express the von\,Neumann entropy on the
given observation level.
Using Eqs. (A.13), (A.14) and (A.17) we obtain
$$
\begin{array}{rl}
M_1={4 \over {Z_G}}\exp \left( {-a+\left| d \right|} \right)  \, , &
\quad
    M_2={4 \over {Z_G}}\exp \left( {-a-\left| d \right|} \right)
\, ,  \\
  M_3={4 \over {Z_G}}\exp \left( {a+\left| b \right|} \right)	\,
, &
\quad
  M_4={4 \over {Z_G}}\exp \left( {a-\left| b \right|} \right).
\end{array}
\eqno({\rm A.20})
$$
The Lagrange multipliers
$\lambda _{kl}$ can be expressed
as functions of the expectation values $\xi _{kl}$:
$$
  \exp \left( a \right)=\left( {{{M_3M_4}
  \over {M_1M_2}}} \right)^{{1 \over 4} }  \, , \quad
\exp \left( {\left| b \right|} \right)=
\left( {{{M_3} \over {M_4}}} \right)^{{1 \over 2}}    \, , \quad
  \exp \left( {\left| d \right|} \right)=
  \left( {{{M_1} \over {M_2}}} \right)^{{1 \over 2}} \, .
\eqno({\rm A.21})
$$
After inserting these expressions into Eq. (A.13) we obtain
for the  partition  function
$$
Z_G={4 \over {\left( {M_1M_2M_3M_4} \right)^{{1 \over 4}}}}
.\eqno({\rm A.22})
$$
When we insert Eqs. (A.18), (A.21) and (A.22) into Eqs.
(A.1), (A.4) and (A.12) then we find  both the entropy
$$
S_G=-\sum\limits_{i=1}^4 {{{M_i} \over 4}}\ln \left( {{{M_i} \over 4}}
\right)
\eqno({\rm A.23})
$$
and the GCDO
$$
\hat{\rho}_G ={1 \over 4}\left( {\matrix{{1+\xi _{zz}}&0&0&{D^*}\cr
0&{1-\xi _{zz}}&{B^*}&0\cr
0&B&{1-\xi _{zz}}&0\cr
D&{}&{}&{1+\xi _{zz}}\cr
}} \right)
\eqno({\rm A.24})
$$
as functions of the expectation values $\xi _{kl}$. Finally,
we can rewrite the reconstructed density operator (A.24)
in terms of the spin operators (see Tab.3).

\subsection*{ A. 2.  Observation level ${\cal O}_H^{(2)}$}

The GCDO on the  ${\cal O}_H^{(2)}$ can be obtained as a result of a reduction
of the observation level ${\cal O}_G^{(2)}$. The difference between these
two observation levels is  that the ${\cal O}_H^{(2)}$ does not contain the operator
$\hat \sigma _z^{(1)}\otimes \hat\sigma _z^{(2)}$, i.e.,
the corresponding mean value is unknown from the measurement.

According to the maximum--entropy principle,
the observation level ${\cal O}_H^{(2)}$ can be obtained from ${\cal O}_G^{(2)}$
by setting the Lagrange multiplier
$\lambda_{zz}$ equal to zero.
With the help of the relation [see Eq.(A.7)]
$$
\lambda_{zz}={{\partial S_G} \over {\partial \xi_{zz}}}
=-{1 \over 4}\ln \left( {{{M_1 M_2} \over {M_3 M_4}}} \right) = 0
\eqno({\rm A.25})
$$
we obtain
$$
M_1M_2=M_3M_4
.\eqno({\rm A.26})
$$
From this equation we find the ``predicted'' mean value of the
operator $\hat \sigma _z^{(1)}\otimes \hat\sigma _z^{(2)}$ (i.e.,
the parameter $t$ in Table 3)
$$
\xi _{zz}={1 \over 4}\left( {\left| D \right|^2
-\left| B \right|^2} \right) \equiv t .
\eqno({\rm A.27})
$$
Taking into account that the parameters $| B|$ and $|D|$ read
$$
\left| B \right|^2=\left( {\xi _{xx}+\xi _{yy}} \right)^2
+\left( {\xi _{xy}-\xi _{yx}} \right)^2  \, , \quad
  \left| D \right|^2=\left( {\xi _{xx}-\xi _{yy}} \right)^2
+\left( {\xi _{xy}+\xi _{yx}} \right)^2 \, ,
\eqno({\rm A.28})
$$
we can express the predicted mean value
$\xi _{zz}$ as a function of the
measured mean values $\xi _{xx}, \,
\xi _{xy},  \, \xi _{yx} \, $ and $ \, \xi _{yy}$:
$$
\xi _{zz}=\left( {\xi _{xy}\xi _{yx}-\xi _{xx}\xi _{yy}} \right) \, .
\eqno({\rm A.29})
$$

When we insert
Eq. (A.27) into Eq. (A.19) we obtain:
$$
  M_1=N_1 N_2  \, , \quad
  M_2=N_3 N_4  \, , \quad
  M_3=N_1 N_3  \, , \quad
  M_4=N_2 N_4\, ,
  \eqno({\rm A.30})
$$
where the parameters $N_i$ are defined as
$$
N_1=1+{1 \over 2}\left( {\left| D \right|+\left| B \right|} \right)
\, , \quad
  N_2=1+{1 \over 2}\left( {\left| D \right|-\left| B \right|} \right)
\, , \\
$$
$$
  N_3=1-{1 \over 2}\left( {\left| D \right|-\left| B \right|} \right)
\, , \quad
  N_4=1-{1 \over 2}\left( {\left| D \right|+\left| B \right|} \right)
\, .
\eqno({\rm A.31})
$$
In addition, from
Eqs. (A.30) and (A.23) we obtain the expression for the
von\,Neumann entropy of the density operator reconstructed on
the observation level ${\cal O}_H^{(2)}$:
$$
S_H=-\sum\limits_{i=1}^4 {{{N_i} \over 2}}
\ln \left( {{{N_i} \over 2}} \right) \, .
\eqno({\rm A.32})
$$
Finally, from Eqs. (A.28) and (A.24) we find the expression
for the GCDO on the observation level  ${\cal O}_H^{(2)}$
(see Table 3):
$$
\begin{array}{rcl}
\hat\rho_{ H}&=&{1 \over 4}{[ \hat I^{(1)}\otimes \hat I^{(2)}+
\left( {\xi _{xy}\xi _{yx}-\xi _{xx}\xi _{yy}} \right)
\hat \sigma _z^{(1)}\otimes \hat \sigma _z^{(2)}} \\
&+& \; {\xi _{xx}\hat \sigma _x^{(1)}\otimes \hat \sigma _x^{(2)}
+\xi _{xy} \hat \sigma _x^{(1)}\otimes \hat \sigma _y^{(2)}}
+\; \xi _{yx}\hat \sigma _y^{(1)}\otimes \hat\sigma _x^{(2)}
+\xi _{yy} \hat\sigma _y^{(1)}\otimes \hat\sigma _y^{(2)}] \, .
\end{array}
\eqno({\rm A.33})
$$



\end{document}